\documentclass[article,onecolumn,superscriptaddress,nofootinbib]{revtex4}
\usepackage{xcolor,color}
\usepackage[a4paper,
            bindingoffset=0.2in,
            left=.7in,
            right=1in,
            top=1in,
            bottom=1in,
            footskip=.25in]{geometry}
\usepackage[utf8]{inputenc}
\usepackage{float}
\usepackage[T1]{fontenc}
\usepackage{latexsym,amssymb,amsmath,amsfonts}
\usepackage{subfigure}
\usepackage{graphicx}
\usepackage{mathrsfs}
\usepackage{float}
\usepackage{hyperref}
\hypersetup{colorlinks=true, linkcolor=magenta, citecolor=green }
\usepackage{enumerate}

\begin{document}

\title{Cosmological eras with the neutrino non-relativistic transition in RTB gravity}
\author{Jonas Pinheiro da Silva\href{https://orcid.org/0000-0001-8456-0096}{\includegraphics[width=10pt]{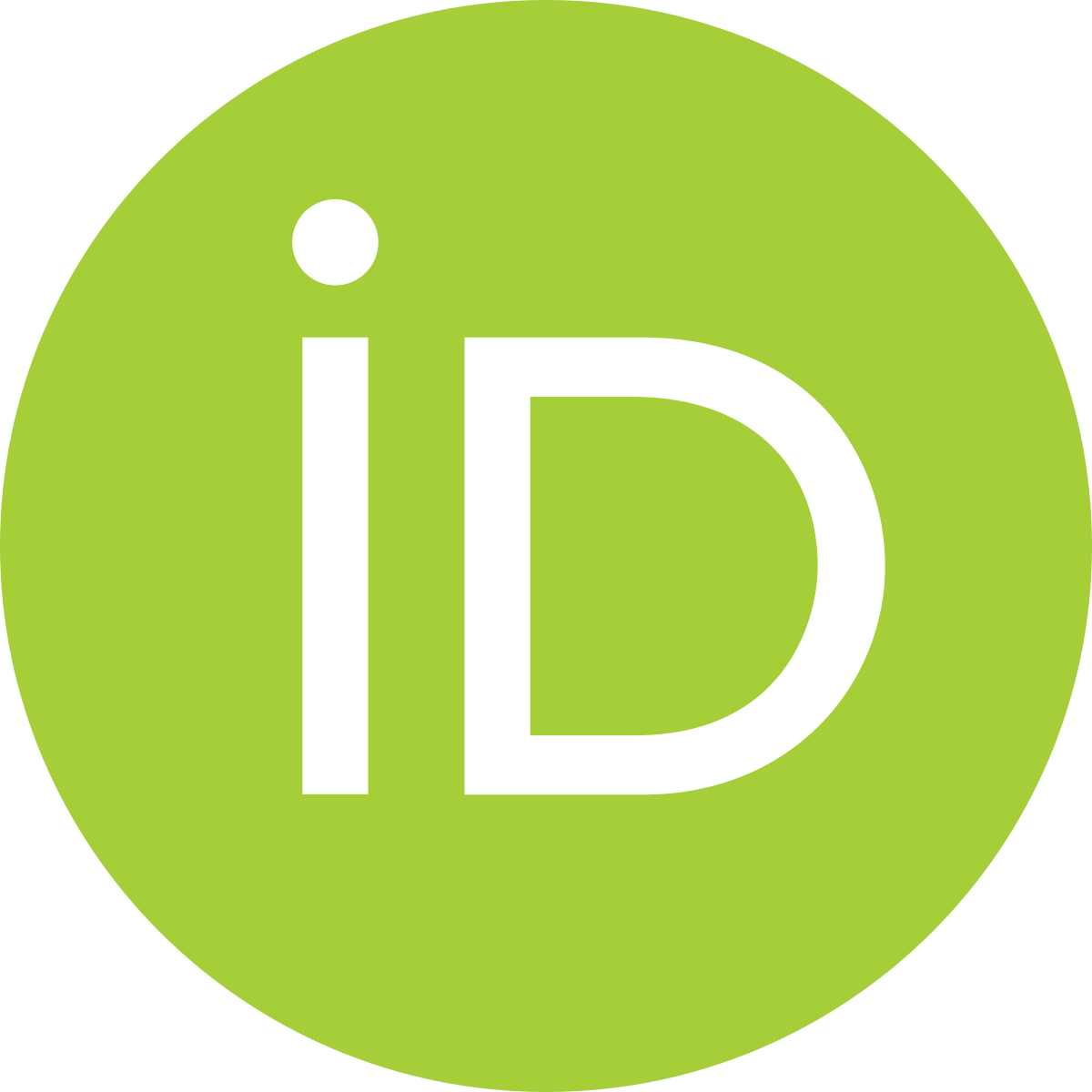}}}
\email{{jonas.j.silva@edu.ufes.br}}
\affiliation{PPGCosmo, CCE, Universidade Federal do Espírito Santo (UFES)\\ Av. Fernando Ferrari, 540, CEP 29.075-910, Vitória, ES, Brasil.}
\affiliation{Instituto de Física, Pontificia Universidad Católica de Valparaíso,
Avenida Brasil 2950, Casilla 4059, Valparaíso, Chile.}
\author{Hermano Velten\href{https://orcid.org/0000-0002-5155-7998 }{\includegraphics[width=10pt]{orcidID.png}}}
\email{hermano.velten@ufop.edu.br}
\affiliation{Departamento de F\'isica, Universidade Federal de Ouro Preto (UFOP), Campus Universit\'ario Morro do Cruzeiro, 35.402-136, Ouro Preto, Brazil}
\date{\today}

\begin{abstract}
    
 Theories based on the Ricci and the trace of the energy-momentum tensor, or the short name Ricci-trace-based (RTB) theories, represent a gravitational approach based on the Ricci scalar $R$, and the trace of the energy-momentum tensor $g^{\mu\nu}T_{\mu\nu} = T$. This theory fits into gravitational models of the type $f(R, T) = R + f(T)$, where $f(T)$ is an arbitrary function of $T$. In this study, we explore a cosmological scenario within the context of RTB models, investigating in detail the cosmological consequences of the coupling between matter and geometry. In order to address this issue, we propose a toy model in which the non-relativistic cosmological neutrino transition plays a role in the cosmic evolution since the effective total energy-momentum tensor trace is affected in this process. We raise questions about the coupling of neutrinos with geometry during this transition, providing a detailed analysis of how RTB gravity deals with this phenomenon and the impact of neutrinos on cosmological dynamics. In summary, we show that the coupling of cosmological neutrinos with $T$ dependent cosmologies is severely challenged.
    \\
    
\end{abstract}

\maketitle
\section{Introduction}

General Relativity (GR) based cosmological scenarios represent the standard way to describe the large scale background expansion of the universe. Given current observational data, an inevitable consequence is the need for an exotic component dubbed dark energy responsible for explaining the present stage of accelerated expansion. This component is usually either in the form of a cosmological constant $\Lambda$ or modelled as a scalar field with a negative equation of state parameter $w_{de}=p_{de}/\rho_{de}$ (by admitting the fluid correspondence), being $p_{de}$ its pressure and  $\rho_{de}$ its energy density.

Due to the lack of direct confirmation on the existence of a fundamental particle/field behind the dark energy phenomena, a widely spread line of investigation concerns the search for viable modified theories of gravity \cite{Tsujikawa:2010zza, Joyce:2016vqv}. Within this approach, GR would be valid only below galactic scales, keeping the regime where it is well tested and some new type of gravitational interaction acts at cosmological scale. The simplest modifications consisting in replacing the Ricci scalar $R$ in the Einstein-Hilbert Lagrangian by generic functions of geometrical quantities $X's$ give rise to the class of $f(R,X's)$ theories. 

We focus on energy-momentum tensor trace $T$ dependent theories of gravity dubbed as $f(R,T)$ models \cite{Harko:2011kv}. This theory has been exhaustively investigated within the astrophysical and cosmological contexts \cite{Harko:2011kv, Mishra:2015era, Sabogal:2024yha, Asghari:2024obf, Devi:2024gcr, Han:2024spn, Sood:2024ufi, Sultana:2024che, Harko:2014gwa, Barrientos:2018cnx, Shabani:2014xvi, Moraes:2019hgx, Jeakel:2023hss, daSilva:2024oov, Velten:2017hhf, Bertini:2023pmp}. It incorporates a coupling between the trace $T$ of the energy-momentum tensor and geometry. Due to the complexity of the coupling structure between matter and geometry, the underlying phenomenology of the coupling between traceless components like radiation and geometry remains a matter of debate. Due to the traceless nature of the radiation component, this implies that theories that encompass the trace of the energy-momentum tensor in geometry do not follow a direct coupling with radiation \cite{Shabani:2014xvi}. We then expect that $f(R,T)$ gravity plays no role in the early universe where the cosmological dynamics is in a good approximation described by a one-fluid component with equation of state parameter $w=p/\rho=1/3$. However, as the universe transits from the radiation domination phase to the matter one, the effective trace contribution becomes a non-vanishing quantity and this switches on modified gravity contributions beyond GR in $T$ dependent modified theories of gravity.   

In this work we aim to explore the cosmological transition from radiation to matter in $f(R,T)$ theories in detail. In doing so, we study the flat Friedmann-Lemaitre-Robertson-Walker (FLRW) background expansion based on $f(R,T)$ gravity sourced by a mixture of radiation, pressureless matter and neutrinos. Cosmological neutrinos suffer a transition from a relativistic to an effective non-relativistic behavior at a redshift $z_t$. This moment depends on the total mass of the three neutrino flavours. This means neutrinos will couple to geometry in $f(R,T)$ gravity after this transition leaving imprints on the background expansion. We discuss in detail the cosmological dynamics in both coupled and decoupled cosmological neutrino scenarios.

In the next section we present the general formalism of $f(R,T)$ gravity and apply it the to cosmological context in the subsequent section. The detailed study of the cosmological evolution in the presence of massive neutrinos in shown in section \ref{SecIV}. We present our final remarks in the final section.

\section{The general fields equations}
The dynamics of gravitational fields, whether in the absence or presence of matter, can be described in terms of either the connection or the metric field. This happens since in differential geometry one can consider the geometric object that connects nearby tangent spaces, the affine connection, as a geometric entity independent (or not) on the metric. The latter case is known as the Palatini or metric-affine formalism. In GR, only one field equation is obtained regardless of the formalism used; whether the metric formalism (only the metric as a dynamic variable) or the metric-affine formalism (metric and connection as dynamic variables). This result holds from the fact that the Riemannian manifold considered is torsion-free. This implies that the Christoffel symbols coincide with the Levi-Civita connection, which is uniquely compatible with the metric and has zero torsion \cite{Olmo:2012yv}. We will keep our focus on the metric formalism for $f(R, T)$ gravity.

\subsection{Metric formalism}

In $f(R, T)$ gravity if one assumes the metric as the unique dynamical variable, as in \cite{Harko:2011kv}, the general action is written as
\begin{eqnarray}\label{action0}
    &&S\left(g_{\mu\nu}, \psi _{m}\right) = \frac{1}{2\kappa ^{2}}\int d^{4}x\sqrt{-g}f(R, T) + \int d ^{4}x \sqrt{-g}\mathcal{L}_{m}(g_{\mu\nu}, \psi _{m}).
\end{eqnarray}
Varying this action with respect to the metric one obtains the following field equations
\begin{eqnarray}\label{metricm}
  &&f_{R}R_{\mu\nu} - \frac{f(R, T)}{2}g_{\mu\nu} + f_{T}\left(T_{\mu\nu} + \Theta _{\mu\nu}\right) + \left(g_{\mu\nu}\Box - \nabla _{\mu}\nabla _{\nu}\right) f_{R}= \kappa ^{2}T_{\mu\nu},
\end{eqnarray}
where $\kappa ^{2} = 8\pi G$ is the coupling term as in GR. $R$ is the Ricci scalar, $T \equiv g^{\mu\nu}T_{\mu\nu}$ is the trace of the energy-momentum tensor, $g$ the metric determinant and $\mathcal{L}_{m}$ the matter lagrangian associated to the fields $\psi _{m}$. We have used the notation $f_{R} = \partial f(R, T)/\partial R$ and $f_{T} = \partial f(R, T)/\partial T$. 
The quantity $\Theta _{\mu\nu}$ has been defined as
\begin{equation}\label{theta}
    \Theta _{\mu\nu} \equiv g^{\mu\nu}\frac{\delta T_{\mu\nu}}{\delta g^{\mu\nu}}.
\end{equation}

The energy-momentum tensor is defined according to
\begin{equation}\label{tem}
    T_{\mu\nu} \equiv - \frac{2}{\sqrt{-g}}\frac{\delta \left(\sqrt{-g}\mathcal{L}_{m}\right)}{\delta g^{\mu\nu}}.
\end{equation}
A detailed derivation of this equation is found in Refs. \cite{Harko:2011kv, Harko:2014gwa}. In section \ref{matter} we will discuss again on its technical aspects.

Given the above expressions and having in mind the Bianchi identities, the covariant derivative of $T_{\mu\nu}$ becomes
\begin{eqnarray}\label{tm}
    &&\nabla _{\mu}T_{\mu\nu} =  \frac{1}{\kappa ^{2}}\biggl[f_{TT}\partial _{\mu}T\left(T_{\mu\nu} + \Theta _{\mu\nu}\right) + f_{T}\nabla _{\mu}\left(T_{\mu\nu} + \Theta _{\mu\nu}\right) - \frac{1}{2}f_{T}\partial _{\mu}T g_{\mu\nu}\biggr].
\end{eqnarray}
It is worth noting that the right hand side of this equation represents a complex combination of geometrical quantities. Only under very specific conditions it vanishes, recovering the conservative aspect of GR. For a discussion on non-conservative theories see Ref. \cite{Velten:2021xxw}.



\subsection{On the matter-geometry coupling}\label{matter}

Since the connection is considered a dynamical variable, both gravity and cosmology are subject to modifications in their dynamical structure. On the other hand, it is possible to verify that some $f(R)$ and $f(T)$ models lead to the same cosmology, regardless of the formalism used. In general, for this concordance to exist, the $f(R)$ model must be linear (up to first order) whereas, with respect to $f(T)$, the models can be non-linear (of higher order in polynomial expansion). In both theories, the coupling with geometry must be minimal; non-minimal coupling between matter and curvature leads to divergences in astrophysical and cosmological solutions for the metric and metric-affine formalism cases \cite{Jeakel:2023hss}. 
In this sense, we are interested in analyzing the dynamics of $f(R, T)$ theory independently on the formalism used in deriving the field equations. In other words, we will work with the minimally coupled scenario i.e., we will consider the model $f(R, T) = R + f(T)$, which we refer to as Ricci-trace-based (RTB). This model is based on the Ricci scalar and an arbitrary function of the trace of the energy-momentum tensor. For this type of model, the contributions from the metric and metric-affine formalism, assuming that the matter is coupled only to the metric, coincide \cite{Afonso:2018hyj}:
\begin{equation}\label{mm}
    \left(g_{\mu\nu}\Box - \nabla _{\mu}\nabla _{\nu}\right)f_{R} = \nabla _{\mu} ^{\Gamma}\left(\sqrt{-g}f_{R}g^{\mu\nu}\right) = 0.
\end{equation}
The first term on the left-hand side of equation \eqref{mm} is the contribution from the Lagrangian density $f(R, T)$ in the metric formalism, equation \eqref{metricm}, and the term on the right-hand side of the first equality is the contribution in the metric-affine formalism  \cite{Barrientos:2018cnx}. Note that in both cases, the coupling of $f(T)$ does not lead to a dependency on the formalism adopted, as both operator and connection equation are associated solely with and by the curvature. The contribution of $f(T)$, in a comprehensive analysis, is a coupling with the geometry. Recently, in order to promote a complete cosmological description, $f(R, T)$ was considered as an interaction theory, where matter and radiation are interacting and both are coupled to geometry \cite{Jeakel:2023hss}. This conjecture was a direct consequence of the way the field equation for this theory has been presented in the literature, more specifically, the presentation of extra terms that arise from the variation of the trace of the energy-momentum tensor. In other words, the equation \eqref{theta} refers to a total energy-momentum tensor. However, the trace of radiation does not contribute to the $f(R, T)$ theory (see \cite{Shabani:2014xvi} for a deeper discussion), since this is a theory whose Lagrangian density is constructed with the coupling with non-relativistic matter. This can be easily demonstrated when we apply the variational principle. By varying the function $f(T)$,
\begin{eqnarray}\label{dt0}
    \delta f(T) = f_{T}\delta T = f_{T}\delta \left(g^{\mu\nu}T_{\mu\nu}\right).
\end{eqnarray}
For the case of a perfect cosmological fluid, the energy-momentum (total) tensor can be described by the sum  of the cosmic components (matter and radiation, respectively),
\begin{equation}
    \mathcal{L}_{m} = L^{m} + L^{r}.
\end{equation}
Consequently, the equation \eqref{tem} is described as follows
\begin{equation}\label{tem1}
    T_{\mu\nu} = T^{m}_{\mu\nu} + T^{r}_{\mu\nu} = - \frac{2}{\sqrt{-g}}\frac{\delta \left(\sqrt{-g}L^{m}\right)}{\delta g^{\mu\nu}} - \frac{2}{\sqrt{-g}}\frac{\delta \left(\sqrt{-g}L^{r}\right)}{\delta g^{\mu\nu}}.
\end{equation}
As a result, since
\begin{equation}\label{trr}
   g^{\mu\nu}T^{r}_{\mu\nu} = 0,
\end{equation}
the variation of $f(T)$ is such that
\begin{equation}
    \frac{\delta f(T)}{\delta g^{\mu\nu}} = f_{T}\left(T^{m}_{\mu\nu} + g^{\mu\nu}\frac{\delta T^{m}_{\mu\nu}}{\delta g^{\mu\nu}}\right).
\end{equation}
Following the last result, only non-relativistic matter couples to the geometry. In the general context, it is easy to show that this statement still holds true for the case with scalar or electromagnetic fields. 

From now on, assuming the model satisfying \eqref{mm} and the conjecture we have assumed, the field equation  \eqref{metricm} is written in the following way
\begin{equation}\label{gr}
    G_{\mu\nu} = \kappa ^{2}T_{\mu\nu} + C_{\mu\nu},
\end{equation}
where we have defined the coupling quantity 
\begin{equation}\label{c}
    C_{\mu\nu} = - f_{T}\left(T^{m} _{\mu\nu} + \Theta ^{m} _{\mu\nu}\right) + \frac{1}{2}f(T)g_{\mu\nu}. 
\end{equation}
Note that \eqref{gr} holds the left hand side of general relativity field equation, but sourced by the sum of the perfect fluid energy-momentum tensor and the $f(T)$ dependent coupling term described in equation \eqref{c}. It is therefore correct the interpretation that this theory can be formally seen as GR plus a geometry dependent exotic fluid.  Therefore, defining 
\begin{equation}\label{newt}
    \mathcal{T}_{\mu\nu} = T_{\mu\nu} + \frac{C_{\mu\nu}}{\kappa ^{2}},
\end{equation}
equation \eqref{gr} is then rewritten in a simplified form as
\begin{equation}\label{newgr}
    G_{\mu\nu} = \kappa ^{2}\mathcal{T}_{\mu\nu}. 
\end{equation}

\section{Cosmological dynamics}
Cosmological solutions for a flat FLRW metric can be obtained from \eqref{newgr}. The tensor \eqref{newt} is explicitly written in therms of the perfect fluid components using signature $(+, -, -, -)$. Given the specific choice $L^{m} = - p_{m}$, then $\Theta ^{m}_{\mu\nu} = -2T^{m}_{\mu\nu} -p_{m}g_{\mu\nu}$. Therefore, 
\begin{eqnarray}\label{totalfluid}
    &&\mathcal{T}_{\mu\nu} = \biggl[\left(\rho _{m} + p_{m}\right)\left(1 + \frac{f_{T}}{\kappa^{2}}\right) + \rho _{r} + p_{r}\biggr]u^{\mu}u_{\nu} + \left(\frac{1}{2\kappa ^{2}}f(T) - p\right)g_{\mu\nu},
\end{eqnarray}
where we have assumed the total energy momentum tensor is composed by a mixture of matter and radiation. This means the total density is $\rho = \rho _{m} + \rho _{r}$ and the total pressure $p = p_{m} + p_{r}$. 

As proposed by \cite{Moraes:2019hgx} and following the analysis promoted in \cite{Jeakel:2023hss, daSilva:2024oov}, let us use the exponential model 
\begin{equation}\label{model}
    f(R, T) = R + \alpha e^{\beta T},
\end{equation}
as the viable $f(R,T)$ approach. Given this $f(R,T)$ model, the Friedmann equation and the four-divergence of the total energy-momentum tensor $\nabla _{\mu}\mathcal{T}^{\mu 0}$ read, respectively
\begin{eqnarray}\label{friedmann}
    3H^{2} = \kappa ^{2}\left(\rho _{m} + \rho _{r}\right) + \alpha e^{\beta \rho _{m}}\left(\beta \rho _{m} + \frac{1}{2}\right),
\end{eqnarray}
and
\begin{eqnarray}\label{cont}
    \dot{\rho}_{m} + \dot{\rho}_{r} + 3H\left(\rho _{m} + \frac{4}{3}\rho _{r}\right) = - \frac{\alpha \beta e^{\beta \rho _{m}}\dot{\rho}_{m}}{\kappa ^{2} + \alpha \beta e^{\beta \rho _{m}}}\left(\beta \rho _{m} + \frac{1}{2}\right). 
\end{eqnarray}

It is possible to set a constraining relation on the Friedmann equation \eqref{friedmann} using $H^{2}(a=1)/H_{0}^{2} = 1$, where the scale factor today (redshift $z=0$) is $a=1$ (see also in \cite{Velten:2017hhf}). With the notation $\Omega_{i0} = \Omega_{i} (z = 0)$ (with $i=m,r$) we obtain

\begin{eqnarray}\label{h0}
    1 = \Omega _{m_{0}} + \Omega _{r_{0}} + \bar{\alpha}e^{\bar{\beta}\Omega _{m_{0}}}\left(\bar{\beta}\Omega _{m_{0}} + \frac{1}{2}\right).
\end{eqnarray}
Rewritting  \eqref{cont} in terms of the fractionary density parameters and using the scale factor as the dynamical variable, where the prime ($^{\prime}$) means derivative with respect to the scale factor,
\begin{eqnarray}\label{ca}
    &&\left(\Omega ^{\prime}_{m} + \Omega ^{\prime}_{r}\right)a + 3\left(\Omega _{m} + \frac{4}{3}\Omega _{r}\right) = - \frac{\bar{\alpha}\bar{\beta}e^{\bar{\beta}\Omega _{m}}\Omega ^{\prime}_{m}a}{1 + \bar{\alpha}\bar{\beta}e^{\bar{\beta}\Omega _{m}}}\left(\bar{\beta}\Omega _{m} + \frac{1}{2}\right).
\end{eqnarray}
The free modified gravity model parameters have been redefined in a dimensionless way as
\begin{equation}
    \bar{\alpha} = \frac{\alpha}{\kappa ^{2}\rho _{0}}\quad {\rm and} \quad \bar{\beta} = \beta \rho _{0}.
\end{equation}

Since radiation is a traceless component it does not couple to geometry (see \eqref{c}). Thus the natural way to proceed is to split \eqref{ca} into two different continuity equations
\begin{equation}\label{densitym}
    \Omega ^{\prime}_{m}a\left[1 + \frac{\bar{\alpha}\bar{\beta}e^{\bar{\beta}\Omega _{m}}}{1 + \bar{\alpha}\bar{\beta}e^{\bar{\beta}\Omega _{m}}}\left(\bar{\beta}\Omega _{m}(a) + \frac{1}{2}\right)\right] + 3\Omega _{m} = 0;
\end{equation}
\begin{equation}\label{rad}
    \Omega ^{\prime}_{r}a + 4\Omega _{r} = 0.
\end{equation}
Of course, the above equation \eqref{rad} lead us to the solution $\Omega _ {r} = \Omega _{r_{0}}a^{-4}$.

After solving numerically \eqref{densitym} for $\Omega_m$ one can insert it into \eqref{friedmann} such that 
 \begin{equation}
     H(a)^{2} =H_{0}^{2}\left[\frac{\Omega_{r_{0}}}{a^4} + \Omega _{m} + \bar{\alpha}e^{\bar{\beta} \Omega _{m}}\left(\bar{\beta}\Omega _{m} + \frac{1}{2}\right)\right].
 \end{equation}
We can identity three different fractionary contributions to the background expansion. They read 
 \begin{equation}\label{fracr}
     \tilde{\Omega} _{r} = \frac{\Omega _{r}}{\Omega _ {r} + \Omega _{m} + \bar{\alpha}e^{\bar{\beta} \Omega _{m}}\left(\bar{\beta}\Omega _{m} + \frac{1}{2}\right)};
 \end{equation}
 \begin{equation}\label{fracm}
     \tilde{\Omega} _{m} = \frac{\Omega _{m}}{\Omega _ {r} + \Omega _{m} + \bar{\alpha}e^{\bar{\beta} \Omega _{m}}\left(\bar{\beta}\Omega _{m} + \frac{1}{2}\right)};
 \end{equation}
 and
  \begin{equation}\label{fracgm}
    \tilde{\Omega} _{gm} = \frac{\bar{\alpha}e^{\bar{\beta} \Omega _{m}}\left(\bar{\beta}\Omega _{m} + \frac{1}{2}\right)}{\Omega _ {r} + \Omega _{m} + \bar{\alpha}e^{\bar{\beta} \Omega _{m}}\left(\bar{\beta}\Omega _{m} + \frac{1}{2}\right)}.
 \end{equation}
The index $gm$ refers to the geometry-matter contribution. 

We address now the question whether the above described dynamics is able to explain available cosmological background data. This analysis has been carefully performed by the same authors in Ref. \cite{Jeakel:2023hss} and there is indeed a small range of the free parameters $\bar{\alpha}$ and $\bar{\beta}$ which can fulfil the minimum requirements to provide a viable cosmological background evolution. We ask the reader to check Fig. 1 of \cite{Jeakel:2023hss}.  Note that if $\bar{\beta}=0$ we recover the flat $\Lambda$CDM model. In this case, the concordance cosmological constant fractionary density value $\Omega_{\Lambda}=0.7$ refers to a value $\bar{\alpha}=1.4$. Then, by keeping $\bar{\beta} \sim 0$ one is also close to the standard cosmology. We have selected three sets of free parameter $(\bar{\alpha}, \bar{\beta})= (1.3, 0.2); (1.1, 0.3); (1., 0.4)$ which are consistent with background data. Values $\bar{\beta} > 0.8$ are not able to fit the data.
 
\begin{figure}[H]
\centering
\includegraphics[width=12
cm]{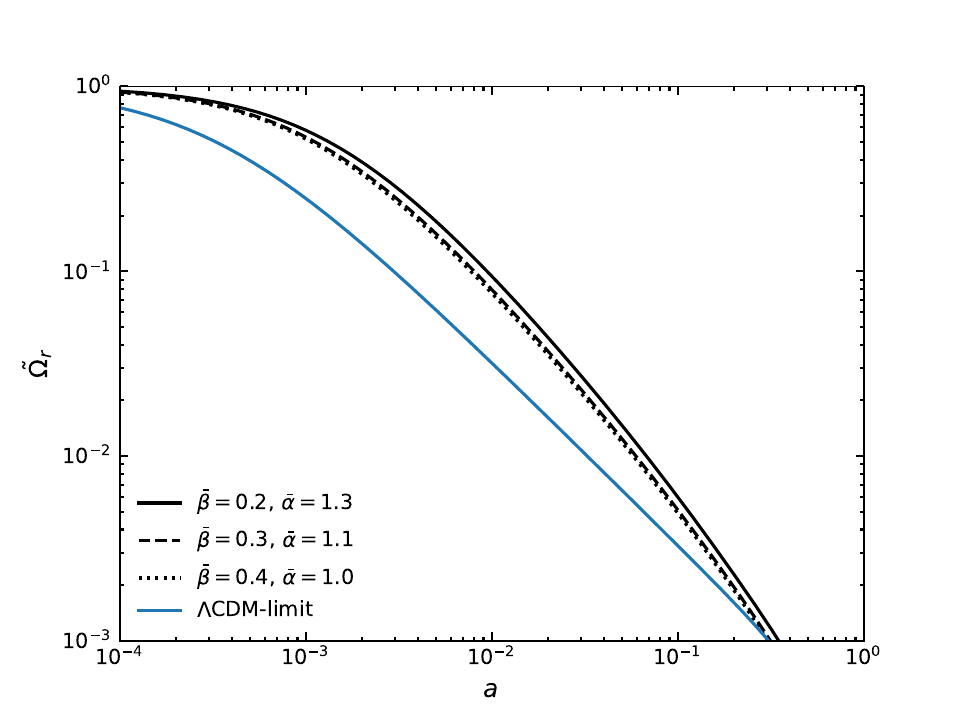}
\caption{Evolution of the fractionary radiation component in terms of the scale factor. In both cases, the radiation density evolves with $a^{-4}$.}
\label{radiation}
\end{figure}

The evolution of \eqref{fracr}, \eqref{fracm} and \eqref{fracgm} is depicted in figures \ref{radiation}, \ref{matter} and \ref{geometry}, respectively. The $\Lambda$CDM limit ($\bar{\alpha} \sim 1.4$ and $\bar{\beta} = 0$) is shown in the blue solid line in all figures.

\begin{figure}[H]
\centering
\includegraphics[width=12cm]{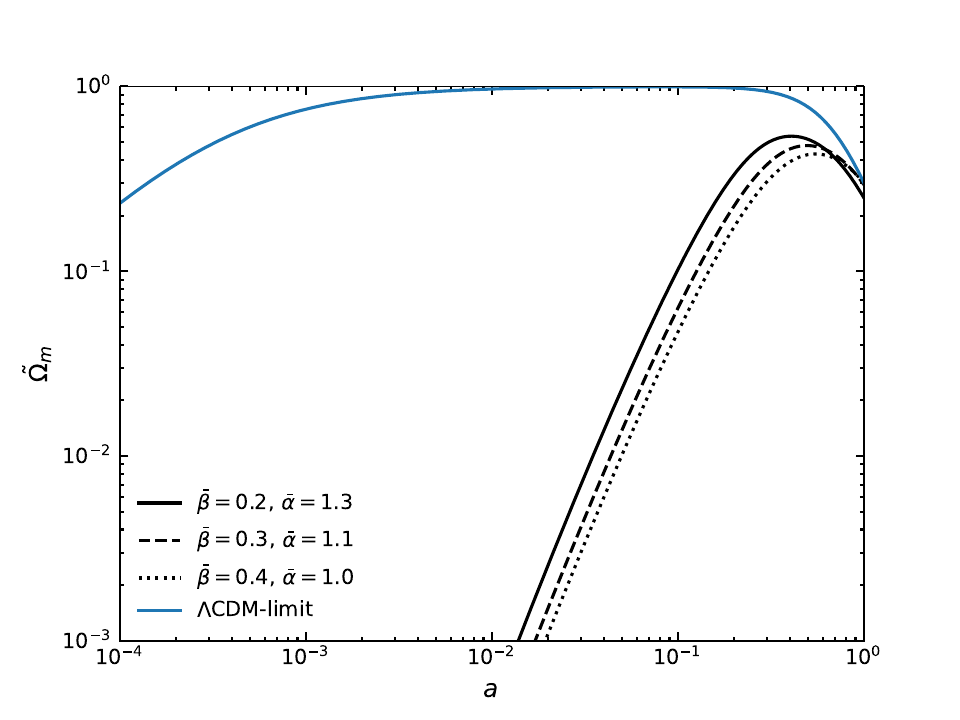}
\caption{The evolution of the fractionary matter component. }
\label{matter}
\end{figure}

\begin{figure}[H]
\centering
\includegraphics[width=12cm]{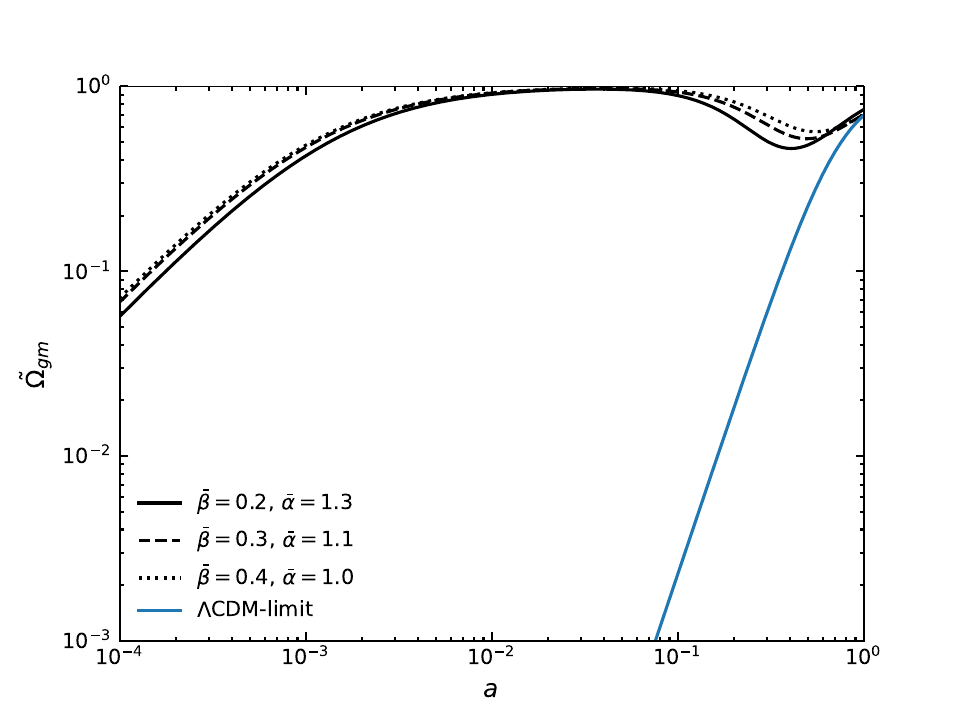}
\caption{The evolution of the fractionary geometry-matter component. In the $\Lambda$CDM limit, the $\Lambda$ contribution evolves according to the solid blue line. The evolution of the geometry-matter component obeys the numerical solution of the differential equation \eqref{ca}. }
\label{geometry}
\end{figure}

From these figures it is clear the universe starts being dominated by the radiation fluid as in the standard cosmology. The radiation domination in the exponential $f(R,T)$ model takes longer than in the $\Lambda$CDM model. The geometry-matter coupled part starts to dominate the dynamics around $z_{eq} \sim 890$. The pure matter component is subdominat. Only around a redshift of order $z\sim 7$ that the matter contribution $\tilde{\Omega}_m$ surpass the contribution of the geometry-matter sector. The geometry-matter component then dominates once again and drives the accelerated expansion observed today. This represents a completely different cosmological scenario when compared to the standard $\Lambda$CDM model. Particularly, the entire process of structure formation should to be rethought in terms of such new dynamical evolution.

\begin{figure}[H]
\centering
\includegraphics[width=8
cm]{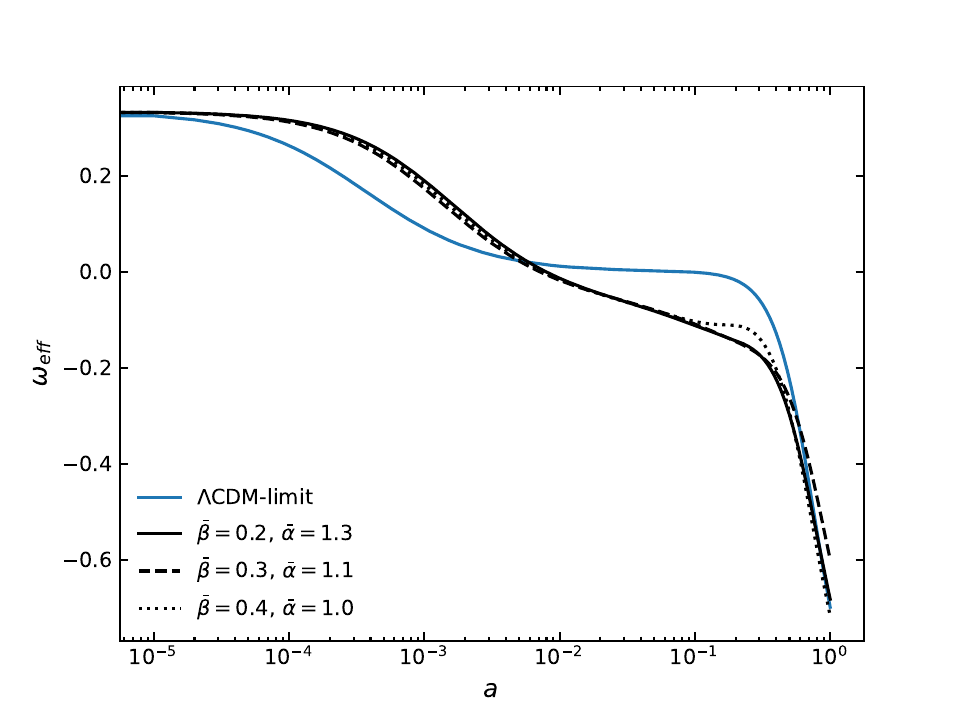}
\includegraphics[width=8
cm]{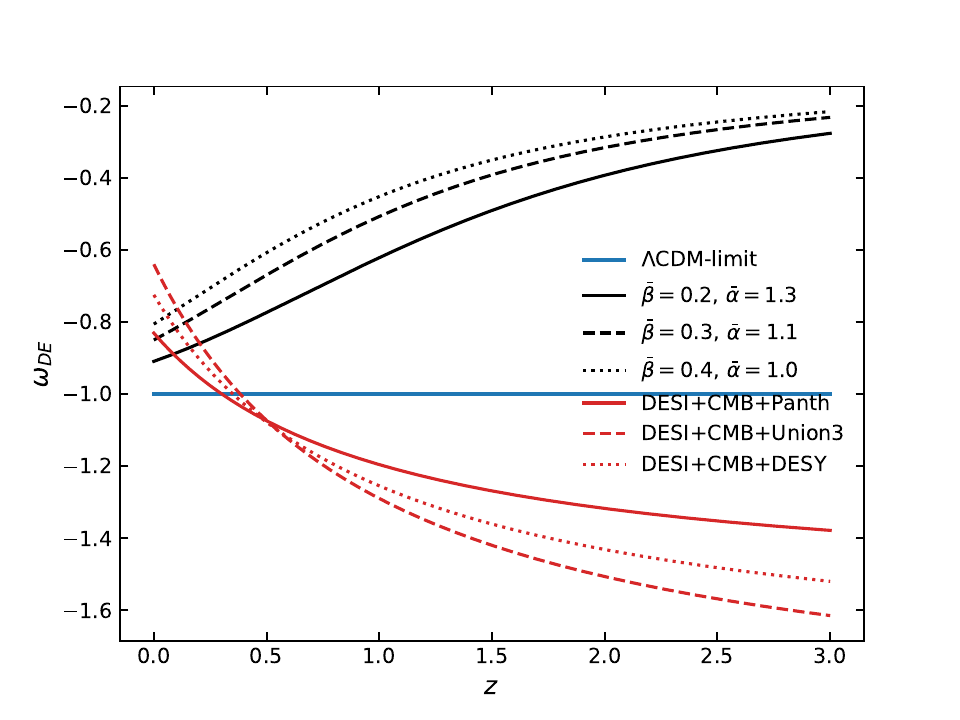}
\caption{Cosmological evolution of the equation of state parameter.}
\label{m6}
\end{figure}

From equation \eqref{friedmann}, we obtain the evolution of the effective equation of state parameter
\begin{equation}
    \omega_{eff} = -1 - \frac{3}{2}\frac{\dot{H}}{H^{2}}, 
\end{equation}
where the equation of state parameter for the geometry-matter sector satisfies
\begin{equation}
   \omega _{gm} \equiv \omega _{DE} = \frac{\rho _{m}f_{T}}{\rho _{m}f_{T} + f(T)/2} -1.
\end{equation}
Here, since the acceleration of the universe is linked to the geometry-matter coupling sector, we denote that the equation of state for this sector is the term commonly used in the literature as ``geometric dark energy'' \cite{Jaime:2013zwa}. The evolution of $\omega_{eff}$ and $\omega_{de}$ is shown in Fig. \ref{m6}. In both figures the solid blue line represents the $\Lambda$CDM equivalent quantity. According to the evolution for $\omega_{eff}$ shown in the black lines in the left panel, after the radiation like behavior ($\omega \approx 1/3$) there is no effective matter dominated phase ($\omega \approx 0$) for a reasonable period. $\omega_{eff}$ cross the line $\omega = 0$ around $z \sim 150$. The right panel of Fig. \ref{m6} shows the evolution of equation of state parameter associated to the dark energy sector. The horizontal blue line refers to the cosmological constant. In red we show the CPL parameterization using the recent values reported by the DESI collaboration \cite{DESI:2024mwx}. The black curves refers to the geometric sector of $f(R,T)$ theories responsible for the accelerated expansion. They behave as quintessence field with $\omega_{de} > -1$, never crossing the phantom dividing line.

\section{Cosmological dynamics in the presence of massive neutrinos}\label{SecIV}

In the last section we have provided the general framework for the cosmological dynamics of $f(R,T)$ gravity using the exponential model \eqref{model} for viable free model parameter values $(\bar{\alpha, \bar{\beta}})$. We explore now the consequence of adding neutrinos to our analysis. This is interesting since neutrinos suffer a non-relativistic transition i.e., they start behaving as relativistic component and then transit to a non-relativistic evolution. Therefore, its energy-momentum trace also transits acquiring a non-vanishing value along this transition and therefore activating the $f(R,T)$ modified gravity sector in this process.

Massive neutrinos can play a fundamental role in explaining the nature of dark matter, being considered in the literature as a good candidate to investigate the stronger interactions between dark matter and the Standard Model. It is discussed that these interactions occur through the neutrino sector, as neutrinos reach observers without being deflected or impeded by other astrophysical phenomena \cite{Datta:2021elq, Arguelles:2022nbl, 10.1063/1.4915588}. After decoupling from the primordial plasma, when the universe is at a temperature of approximately $1MeV$, neutrinos begin to propagate freely through space-time with a comoving total energy associated with their $i-th$ state, $\epsilon_{i}$ \cite{Lesgourgues:2014zoa, Kamalinejad:2022yyl}
\begin{equation}
    \epsilon _{i} = \sqrt{q_{i}^{2} + a^{2}m_{\nu , i}^{2}},
\end{equation}
where $q_{i}$ is the comoving moment of the $i$ state which decays as $a^{-1}$  and $m_{\nu , i}$ is its rest mass.
Depending on the mass of the neutrino, as the universe expands, the comoving total energy of the neutrino is dominated by its rest mass, making them non-relativistic. That is, neutrinos start to behave like matter rather than radiation. The moment of this transition, in terms of the scale factor, is given by
\begin{equation}
    a_{T, i} ^{-1} = 1890\left(\frac{m_{\nu, i}}{1eV}\right).
\end{equation}
According to Ref. \cite{Kamalinejad:2022yyl}, this can be translated into an effective equations of state parameter such as
\begin{equation}\label{estneu}
        \omega _{\nu} (a) = \frac{1}{3\left[1 + (a/a_{T})^{2}\right]}.
\end{equation}
Note that from now on, we neglect the subscript $i$ that accompanied the scale factor at the moment of transition. This is due to the fact that equation \eqref{estneu} is an approximation, since the redshift of the transition occurs when the kinetic energy of the neutrinos is equivalent to the rest mass, which implies that the neutrinos are not completely non-relativistic at the moment of transition (see \cite{Kamalinejad:2022yyl} for more details). Based on the recent constraints reported for the standard cosmological model and quintessence (see \cite{Yarahmadi:2023uqj}), we will work with specific choices of the total neutrino mass \cite{Lesgourgues:2014zoa}. Assuming $m_{\nu, i} < 1eV$, the neutrino transition should occur when the universe is matter-dominated.

In the context of $f(R, T)$ gravity, the implementation of neutrinos can be done, on one hand, at the Lagrangian level resulting in neutrinos always being decoupled to geometry. This means neutrinos complement the radiation component with photons, $L^{r} = - p_{p} - p_{\nu}$, where $p_p$ is the pressure exerted by the photons and $p_{\nu}$ the pressure of the massive neutrinos. On the other hand, the inclusion of neutrino sector can also occur at the level of the material content of the cosmological fluid, providing us with a new structure for the trace of the energy-momentum tensor (see equation \eqref{tneu} below). The characteristic equation for the evolution of neutrinos is given by
\begin{equation}
    \Omega _{\nu}(a) = \Omega _{\nu _{0}}a^{-3}\left[\frac{1 + (a_{T}/a)^{2}}{1 + a^{2}_{T}}\right]^{1/2},
\end{equation}
where $\Omega _{\nu _{0}} = m_{\nu, i}/h^{2}93.14eV$. As mentioned earlier, massive neutrinos still exert pressure even after the transition and, for $f(R, T)$ theory, this is an important signature along the non-relativistic transition. When neutrinos cease to be relativistic, equation \eqref{estneu} suggests that neutrinos, even with non-zero pressure, could couple to geometry, unlike photons. In other words, there are two possible conjectures for the analysis of neutrinos in $f(R, T)$. If we initially implement neutrinos in the relativistic Lagrangian, neglecting their mass, we will have only baryons and dark matter coupled to geometry. Neutrinos would always be free and would not couple to geometry. On the other hand, if we explicitly state the form of the total Lagrangian, decoupling the neutrino sector, the trace of the energy-momentum tensor is then described by 
\begin{equation}\label{tneu}
    T = \rho _{m} + \rho _{\nu}\left[1 - \frac{1}{1 + (a/a_{t})^{2}}\right],
\end{equation}
and not only by  $T = \rho _{m}$. This last approach indicates neutrinos would be always coupled to geometry.

When coupled to geometry, the cosmological evolution of neutrinos is described in $f(R,T)$ gravity obeying the non-conservative nature of $f(R, T)$ theory, $\nabla _{\mu}T^{\mu 0}_{\nu} \neq 0$, leading us to
\begin{eqnarray}\label{neutrinoevolution}
   && \Omega ^{\prime}_{\nu c}a + 3\Omega _{\nu c}\left(1 + \omega _{\nu}\right) = - \frac{a \bar{\alpha}\bar{\beta}e^{\bar{\beta}\left[\Omega _{m} + \Omega _{\nu c}\left(1 - 3\omega _{\nu}\right)\right]}}{1 + \bar{\alpha}\bar{\beta}e^{\bar{\beta}\left[\Omega _{m} + \Omega _{\nu c}\left(1 - 3\omega _{\nu}\right)\right]}}\biggl\{\bar{\beta}\bigl[\Omega ^{\prime}_{m}\Omega _{\nu c}(1 + \omega _{\nu})\bigr] + \bar{\beta}\bigl[\Omega ^{\prime}_{\nu c}(1 - 3\omega _{\nu}) - \nonumber \\
   && - 3\Omega _{\nu c}\omega ^{\prime}_{\nu}\bigr]\bigl[\Omega _{m} + \Omega _{\nu c}(1 + \omega _{\nu})\bigr] + \Omega ^{\prime}_{\nu c}\omega _{\nu} + \Omega _{\nu c}\omega ^{\prime}_{\nu} +\frac{1}{2}\bigl[\Omega ^{\prime}_{\nu c}(1 - 3\omega _{\nu}) -3\Omega _{\nu c}(a)\omega^{\prime}_{\nu}\bigr]\biggr\}. 
\end{eqnarray}
The subscript $\nu c$ refers to the case in which neutrinos are always coupled to geometry.

 We are interested in observing the evolution of the trace of the modified energy-momentum tensor given by \eqref{totalfluid}. In this sense, the following definitions refer to the cases in which neutrinos are free \eqref{tunc} and coupled \eqref{tc} to geometry, respectively,
 \begin{eqnarray}\label{tunc}
     \mathcal{T} = \rho _{0}\left\{\Omega _{m}\left(1 + \bar{\alpha}\bar{\beta}e^{\bar{\beta}\Omega _{m}}\right) + 2\bar{\alpha}e^{\bar{\beta}\Omega _{m}} + \Omega _{\nu}\left[1 - \frac{1}{1 + (a/a_{t})^{2}}\right]\right\},
 \end{eqnarray}
 \begin{eqnarray}\label{tc}
     \mathcal{T}_{c} = \rho _{0}\left\{\Omega _{m} + \Omega _{\nu c}\left[1 - \frac{1}{1 + (a/a_{t})^{2}}\right] + \bar{\alpha}e^{\bar{\beta}[\Omega _{m} + \Omega _{\nu c}(1 - 3\omega _{\nu})]}\left[\bar{\beta}\Omega _{m} + \bar{\beta}\Omega _{\nu c}(1 + \omega _{\nu}) + 2\right]\right\}.
 \end{eqnarray}
 Relations \eqref{tunc} and \eqref{tc} depend on the conservation equation \eqref{neutrinoevolution}. However, in the the first case, the decoupled one, we set the right hand side of \eqref{neutrinoevolution} being zero, as it pertains to the case where the neutrinos are free (not coupled to geometry).

The Hubble expansion taking into account matter, radiation and neutrinos is therefore written according to the expression
\begin{eqnarray}\label{h1}
    &&\frac{H^{2}_{c}(a)}{H^{2}_{0}} = \Omega _{m} + \Omega _{r} + \Omega _{\nu c
} + \bar{\alpha}e^{\bar{\beta}\left[\Omega _{m} + \Omega _{\nu c}\left(1 - 3\omega _{\nu}\right)\right]}\biggl[\bar{\beta}\Omega _{m} + \bar{\beta}\Omega _{\nu c} (1 + \omega _{\nu}) + \frac{1}{2}\biggr],
\end{eqnarray}
for the coupled case, whereas for the decoupled case one has
\begin{eqnarray}
    &&\frac{H^{2}(a)}{H^{2}_{0}} = \Omega _{m} + \Omega _{r} + \Omega _{\nu} + \bar{\alpha}e^{\bar{\beta}\Omega _{m}}\biggl[\bar{\beta}\Omega _{m} + \frac{1}{2}\biggr].
\end{eqnarray}
 
\begin{figure}[H]
\centering
\includegraphics[width=8
cm]{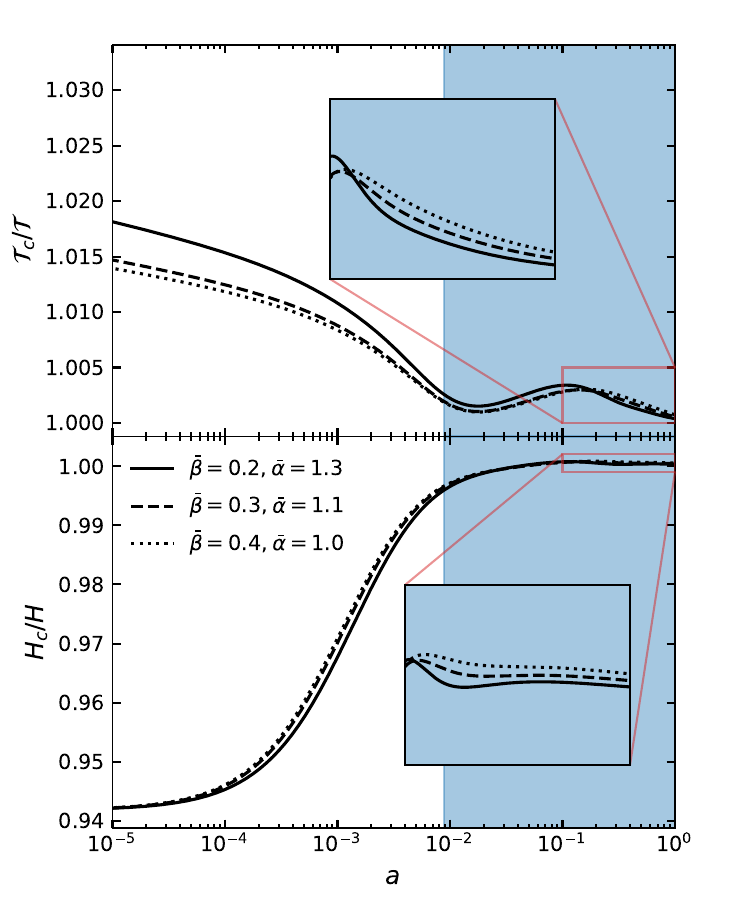}
\includegraphics[width=8
cm]{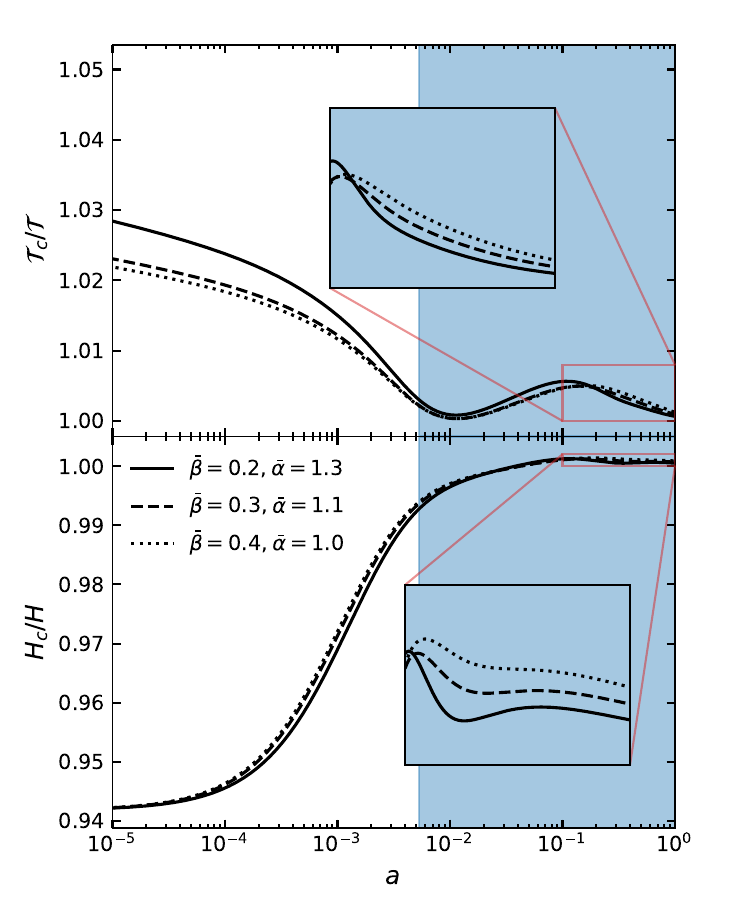}
\caption{Evolution of the ratio $\mathcal{T}_c/\mathcal{T}$ and $H_c/H$ adopting the sum of the neutrinos species $m_{\nu} = 0.06$ in the left panel and $m_{\nu} = 0.1$ in the right panel. }
\label{m5}
\end{figure}

 Based on equation \eqref{estneu}, we calculate the ratio $\mathcal{T}_{c}/\mathcal{T}$ in order to observe the discrepancy between the coupled and decoupled  cases. The results can be seen in Figure \ref{m5}.

\begin{figure}[H]
\centering
\includegraphics[width=12cm]{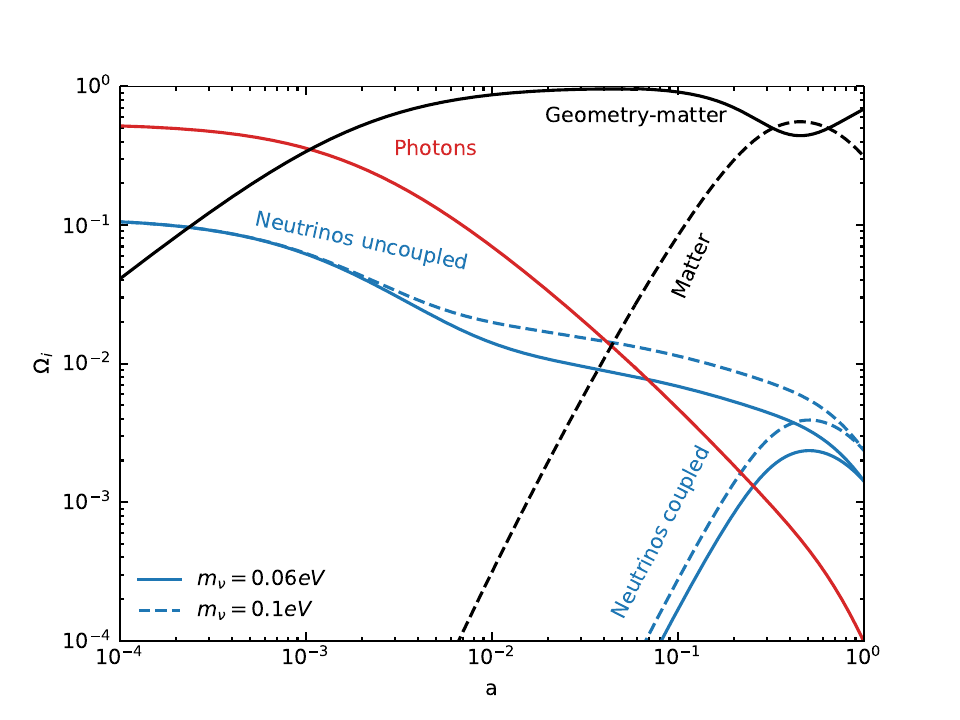}
\caption{Total fractional density of all components for the exponential model. The evolution of decoupled neutrinos is obtained when the equation \eqref{neutrinoevolution} is conserved, preserving the left-hand side of the equation. We fixed $\bar{\alpha} = 1.1$ and $\bar{\beta} = 0.3$. }
\label{figneu}
\end{figure}

 In Figure \ref{figneu}, it is possible to observe the evolution of all components of the modified cosmological fluid described in \eqref{totalfluid}, adding the neutrino sector to the sum of the components (matter and radiation). The sum of all neutrinos species is also shown. The blue shaded region represents the moment after the non-relativistic transition. Note that the non-relativistic neutrino transition occurs when the universe is dominated by the geometry-matter component. As expected, the neutrino contribution to the total expansion is not relevant. In the coupled case it is even less important, suggesting that in $f(R,T)$ theories the neutrino sector would leave no detectable imprint on cosmological observables. This a quite remarkable aspect of the model studied here.

\section{Conclusions}

We have explored the cosmological evolution in $f(R,T)$ modified gravity theories by considering the exponential model presented in \eqref{model}. We have clearly identified the contribution from each cosmic component in this model, i.e., a radiation fluid, a pressureless matter fluid and the geometry-matter component. The latter becomes the dominant one at redshifts of order $\sim 890$. There is no a pure Einstein-de Sitter (matter dominated) expansion as in the standard cosmological scenario. The matter component turns to dominate the expansion around a redshift $\sim 3$ but it is surpassed by the geometry-matter component later on around $z\sim 1.9$. Therefore, one can identify four distinct cosmological era instead of the three ones as in the standard cosmological model.   

We have also paid attention to the role played by neutrinos in this kind of modified gravity scenario. Since the modified gravity sector depends on the trace of the energy momentum tensor, we have analysed how the non-relativistic transition of cosmological neutrinos would affect the effective energy-momentum trace. We identified two different situations: neutrinos are either coupled or decoupled to the geometrical sector. For the decoupled case, their fractionary energy density parameter behaves quite differently from the standard scenario. As one can see in Fig. \ref{figneu}, with the coupling of neutrinos to geometry they present an almost vanishing contribution at high redshifts, orders of magnitude smaller then the standard decoupled case. This means neutrinos would barely influence the early universe and their signatures via e.g., the early integrated Sachs-Wolfe effect would be undetectable.

\section*{Acknowledgments}

We thank CNPq, CAPES, FAPEMIG and FAPES for financial support.
	
\end{document}